\documentclass[aps,prd,preprint]{revtex4}

\usepackage[encapsulated]{CJK}
\usepackage{lipsum}

\usepackage{graphicx}
\usepackage{graphicx}
\usepackage{hyperref}
\usepackage{amsmath}
\usepackage{url}
\usepackage{colordvi}

\usepackage{times}
\begin{document}

\begin{CJK}{UTF8}{bsmi}

\title{On the extraction of the intrinsic light-quark sea in the 
proton}

\author{Wen-Chen Chang (章文箴)}
\affiliation{Institute of Physics, Academia Sinica, Taipei 11529, Taiwan}
\author{Jen-Chieh Peng (彭仁傑)}
\affiliation{Department of Physics, University of Illinois at Urbana-Champaign, Urbana, Illinois 61801, USA}

\begin{abstract}
The HERMES collaboration recently reported a reevaluation of the
strange-quark parton distribution, $S(x)$, based on kaon production in
semi-inclusive deep-inelastic scattering. Two distinct results on
$S(x)$ at the $x> 0.1$ region, one with a sizable magnitude and
another with a vanishing content, were reported. We show that the
latter result is due to a particular assumption adopted in the
analysis. The impact of the new HERMES $S(x)$ result on the extraction
of intrinsic light-quark sea in the proton is discussed. Given the
large uncertainty in the kaon fragmentation function, we find that the
latest HERMES data do not exclude the existence of a significant
intrinsic strange-quark sea in the proton. The $x$ dependence of the
$(s+\bar s)/(\bar u+ \bar d)$ ratio is also in qualitative agreement
with the presence of intrinsic strange-quark sea.
\end{abstract}


\maketitle
\end{CJK}

\section{Introduction}

The large magnitude of the coupling constant, $\alpha_s$, in strong
interaction implies that sea quarks, like valence quarks and gluons,
represent an integral part of the nucleon's structure.  Unlike the
valence quarks in the nucleons, which are restricted to the $u$ and
$d$, the sea quarks can have any quark flavors. This leads to a
potentially rich flavor structure for the nucleon sea and could offer
new insights on the nucleon structure. While decades of experimental
and theoretical work has focused on the valence quark distributions,
many important properties of the sea quarks, including their flavor,
spin, and momentum dependencies, remain to be better determined.

A major surprise in the flavor structure of the nucleon sea was found
when deep-inelastic scattering (DIS) and Drell-Yan experiments showed
that the $\bar u$ and $\bar d$ in the proton have strikingly different
Bjorken-$x$
dependence~\cite{tony,kumano,vogt,garvey,peng_qiu,chang_peng}. Theoretical
models which can explain this flavor asymmetry also have specific
predictions on other aspects of spin and flavor structures of sea
quarks. For example, the $s(x)$ and $\bar s(x)$ distributions are
predicted to be different in, e.g., the meson cloud
model~\cite{signal87,holtmann96}, the statistical
model~\cite{soffer,bhalerao}, and the chiral soliton
model~\cite{wakamatsu03}. It is also interesting to investigate how
the flavor asymmetry between $\bar u$ and $\bar d$ is extended to the
SU(3) case when the $s$ and $\bar s$ seas are included.

Our knowledge on the strange quark contents in the nucleons comes
primarily from neutrino DIS and charged-lepton semi-inclusive DIS
(SIDIS) experiments. From neutrino DIS, the momentum fraction carried
by $s + \bar s$, integrated over the measured $x$ range, is found to
be roughly half of that carried by the lighter $\bar u + \bar d$
quarks~\cite{CCFR95,NOMAD2013}, reflecting a broken SU(3) symmetry for
the proton's sea.  In 2008, the HERMES collaboration reported a
determination of $x(s(x) + \bar s(x))$ over the range of $0.02 < x <
0.5$ at ${\rm Q}^2 = 2.5$ GeV$^2$ from their measurement of SIDIS for
charged-kaon production on a deuteron target with $e^\pm$
beam~\cite{hermes08}. The invariant mass of the photon-nucleon
  system $W$ is required to be greater than $\sqrt{10}$ GeV. The
HERMES result shows an intriguing feature that $x(s(x) + \bar s(x))$
for $x<0.1$ rises rapidly with decreasing $x$, becoming comparable to
$x(\bar u(x) + \bar d(x))$ from the CTEQ6L parton distribution
function (PDF)~\cite{CTEQ6L} at $x<0.05$. The Fermilab E866 Drell-Yan
experiment also shows that the $\bar d(x) / \bar u(x)$ ratio
approaches unity at the lowest value of $x$
($\sim$0.02)~\cite{E866}. These results suggest the presence of an
SU(3) flavor symmetric proton sea in the small-$x$ region.  Recently,
the ATLAS collaboration determined the strange-to-down sea-quark ratio
$r_s$ ($\equiv (s + \bar s)/2 \bar d)$) to be 1.00$^{+0.25}_{-0.28}$
at $x=0.023$ and ${\rm Q}^2 = 1.9$ GeV$^2$ from an analysis of
inclusive $W$ and $Z$ boson production in $pp$ collisions at 7
TeV~\cite{Atlas_WZ}. Furthermore, the ATLAS collaboration determined
the same ratio $r_s$ to be $0.96^{+0.26}_{-0.30}$ at ${\rm Q}^2$=1.9
GeV$^2$ from their measurement of the associated $W+c$ production at
LHC~\cite{Atlas_Wc}. These results strongly suggests an
SU(3)-symmetric light-quark sea, i.e., $\bar u = \bar d = \bar s$, in
the small-$x$ region. Results from HERMES, ATLAS, as well as the
earlier neutrino results, imply a strong $x$ dependence for the
$[s(x)+\bar s(x)]/[\bar u(x) + \bar d(x)]$ ratio.

Another intriguing aspect of the nucleon sea is the concept of
``intrinsic" sea, suggested by Brodsky, Hoyer, Peterson, and Sakai
(BHPS)~\cite{brodsky80} to explain the enhanced production rates for
charmed hadrons in the forward rapidity region. The $c \bar c$
component in the $|uudc \bar c\rangle$ is called the ``intrinsic" sea
in order to distinguish it from the conventional ``extrinsic" sea
originating from the $g \to c \bar c$~ QCD process. The intrinsic sea
is predicted to have a valence-like momentum distribution peaking at
relatively large $x$. This is in contrast to the extrinsic sea, which
dominates in the small-$x$ region due to gluon splitting. Recently,
the concept of intrinsic charm was generalized to the light-quark
sector~\cite{chang11}. Since the probability for the $|uud\mathcal{Q}
\bar{\mathcal{Q}}\rangle$ Fock state is expected to be roughly
proportional to $1/m^2_\mathcal{Q}$, where $m_\mathcal{Q}$ is the mass
of quark $\mathcal{Q}$, these light-quark intrinsic seas should be
more abundant than the intrinsic charm quark. Using the kaon SIDIS
data from HERMES on $s(x) + \bar s(x)$~\cite{hermes08}, the E866
Drell-Yan data on $\bar d(x) - \bar u(x)$~\cite{E866}, and the CTEQ6.6
PDF~\cite{CTEQ6.6}, it was shown that the probabilities for the
$|uudu\bar u\rangle$, $|uudd\bar d\rangle$, and $|uuds\bar s\rangle$
Fock states have been extracted, supporting the existence of the
intrinsic light-quark sea~\cite{chang11}.

Recently, the HERMES collaboration reported the latest result on
charged-kaon multiplicities where a multi-dimensional unfolding
procedure was performed~\cite{hermes13}. This led to a reevaluation of
the strange-quark distributions, $S(x) \equiv s(x) + \bar s(x)$
~\cite{hermes14}. Depending on the kaon fragmentation function adopted
in the analysis, two distinct results, corresponding to significantly
different $S(x)$ at the $x > 0.1$ region, were obtained by the HERMES
collaboration~\cite{hermes14}. Since the HERMES result on $S(x)$ is a
crucial input for the extraction of the intrinsic light-quark sea, we
have extended our previous analysis~\cite{chang11} to take into
account the latest HERMES results. In this paper, we first discuss the
uncertainties in the extraction of $S(x)$ associated with the
uncertainty in the kaon fragmentation functions which are still poorly
known. We then recapitulate the procedure to extract the intrinsic
sea in the BHPS model and present the updated results on the
extraction of the intrinsic light-quark sea using several different
assumptions for the kaon fragmentation functions. The $x$ dependence
of the $(s(x) + \bar s(x))/(\bar u(x) + \bar d(x))$ ratio is also
discussed in the context of the extrinsic and intrinsic seas.

\section{Extraction of S(x) from HERMES kaon SIDIS data}

\begin{figure}[htb]
\centering
\includegraphics[width=0.9\textwidth]{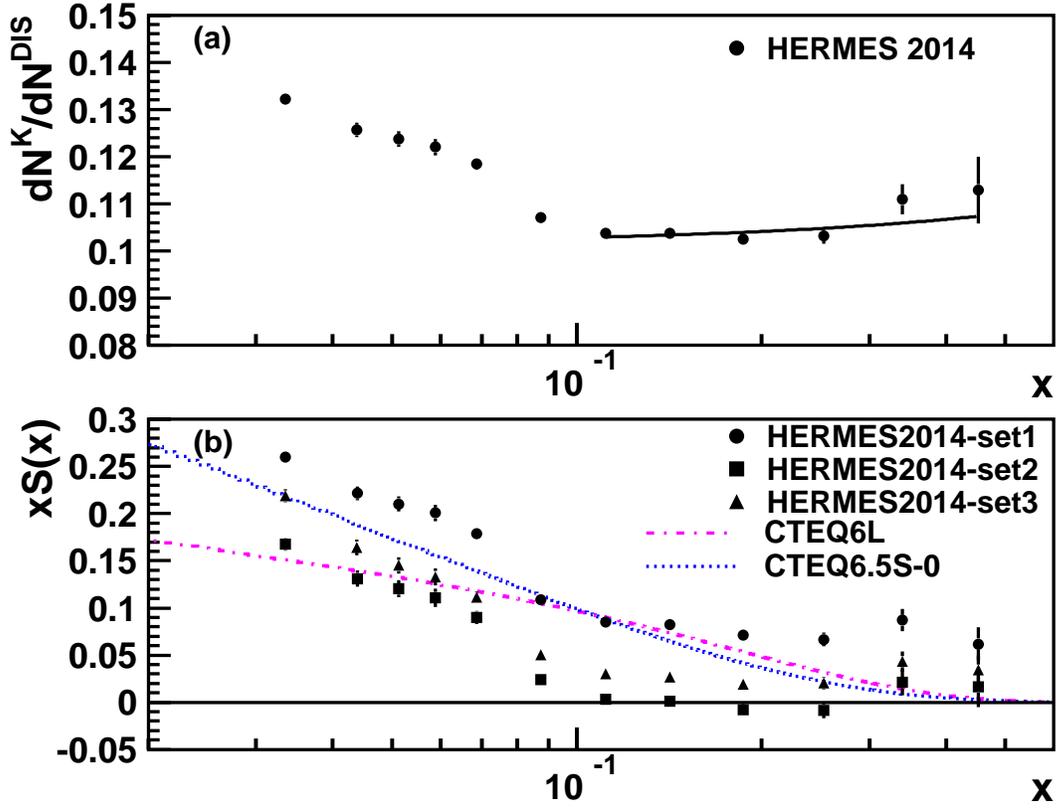}
\caption{(a) The fit to the ${dN^K}/{dN^{DIS}}(x,{\rm Q}^2)$ HERMES
  data~\cite{hermes14} (solid circles) for the determination of an
  effective value of $\int D_Q^K(z,{\rm Q}^2)dz$. (b) The strange
  parton distribution $xS(x)$ determined from the measured
  charged-kaon multiplicities shown in the Fig.~2 (HERMES2014-set1)
  and Fig.~4 (HERMES2014-set2) of Ref.~\cite{hermes14}. Also shown 
is the $xS(x)$ distribution of HERMES2014-set3, described in
the text. For clarity, only statistics errors are shown. Systematic uncertainties are available in Ref.~\cite{hermes14}.}
\label{fig1}
\end{figure}

The HERMES collaboration extracted $S(x)$ from the spin-averaged kaon
multiplicity, ${dN^K}(x,{\rm Q}^2)/{dN^{DIS}}(x,{\rm Q}^2)$, measured 
with 27.6 GeV positrons or electrons scattered off a deuterium target 
in the DIS region~\cite{hermes13}. The HERMES ${dN^K}(x,{\rm Q}^2) / 
{dN^{DIS}}(x,{\rm Q}^2)$ data are shown in Fig.~\ref{fig1}(a). 
For the isoscalar deuteron nucleus, the kaon multiplicity is expressed
in leading order
as follows~\cite{hermes08,hermes14}:
\\
\begin{equation}
\frac{dN^K(x,{\rm Q}^2)}{dN^{DIS}(x,{\rm Q}^2)}=\frac{Q(x,{\rm Q}^2) \int_{0.2}^{0.8} 
D_Q^K (z,{\rm Q}^2)dz +
  S(x,{\rm Q}^2) \int_{0.2}^{0.8} D_S^K (z,{\rm Q}^2)dz}{5Q(x,{\rm Q}^2)+2S(x,{\rm Q}^2)}
\label{eq1}
\end{equation}
\\
where $S(x,{\rm Q}^2) \equiv s(x,{\rm Q}^2)+\bar s(x,{\rm Q}^2)$ and 
$Q(x,{\rm Q}^2) \equiv u(x,{\rm Q}^2) + \bar u(x,{\rm Q}^2)
+ d(x,{\rm Q}^2) + \bar d(x,{\rm Q}^2)$. The $D_S^K(z,{\rm Q}^2)$ 
and $D_Q^K(z,{\rm Q}^2)$ are their corresponding
fragmentation functions for hadronizing into charged kaons.
The values of 0.2 and 0.8 are the lower and upper limits of the
variable $z=E_K/\nu$, where $\nu$ and $E_K$ are the energies
of the virtual photon and the kaon in the target rest frame, respectively.
Eq.~(\ref{eq1}) can be rearranged as 
\\
\begin{equation}
S(x,{\rm Q}^2) = \frac{Q(x,{\rm Q}^2)
[5(\frac{dN^K(x,{\rm Q}^2)}{dN^{DIS}(x,{\rm Q}^2)})
- \int_{0.2}^{0.8} D_Q^K (z,{\rm Q}^2)dz]}
{\int_{0.2}^{0.8} D_S^K (z,{\rm Q}^2)dz
- 2(\frac{dN^K(x,{\rm Q}^2)}{dN^{DIS}(x,{\rm Q}^2)})}.
\label{eq2}
\end{equation}
\\

Eq.~(\ref{eq2}) shows that $S(x,{\rm Q}^2)$ can be directly evaluated
by using the HERMES kaon multiplicity data, the values of $Q(x,{\rm
  Q}^2)$ from recent PDFs, and $D_Q^K (z,{\rm Q}^2)$, $D_S^K (z,{\rm
  Q}^2)$ from the latest parametrization of kaon fragmentation
functions (FF). The result of $xS(x)$ for an evaluation using the
CTEQ6L PDF~\cite{CTEQ6L} and the DSS FF~\cite{DSS} are shown in Fig. 2
of Ref.~\cite{hermes14} and as ``HERMES2014-set1'' in
Fig.~\ref{fig1}(b) (statistical errors only). For the
  $0.03 < x < 0.2$ range, the extracted $xS(x)$ values are much larger
  than what are predicted by CTEQ6L PDF but closer to those of
  CTEQ6.5S-0~\cite{CTEQ6.5s}, a reference PDF set with unconstrained
  $xS(x)$ shape. For the three largest $x$ points at $x > 0.2$, the
extracted $xS(x)$ values are larger than the PDF values, although the
differences are comparable to the large systematic uncertainties
estimated by the HERMES collaboration~\cite{hermes14}.


An alternative approach was adopted by the HERMES collaboration where the
integral $\int_{0.2}^{0.8} D_Q^K (z,{\rm Q}^2)dz$ was estimated 
using their kaon multiplicity
data. Eq.~(\ref{eq1}) could be rearranged as
\\
\begin{equation}
\int_{0.2}^{0.8} D_Q^K(z,{\rm Q}^2)dz  = 5\frac{dN^K(x,{\rm Q}^2)}
{dN^{DIS}(x,{\rm Q}^2)}  -  \frac{S(x,{\rm Q}^2)}
{Q(x,{\rm Q}^2)}[\int_{0.2}^{0.8} D_S^K(z,{\rm Q}^2)dz 
- 2\frac{dN^K(x,{\rm Q}^2)}{dN^{DIS}(x,{\rm Q}^2)}].
\label{eq3}
\end{equation}
\\
When $S(x,{\rm Q}^2)/Q(x,{\rm Q}^2)$ is sufficiently small, the 
second term on the
RHS of Eq.~(\ref{eq3}) can be neglected with respect to 
the first term, and Eq.~(\ref{eq3}) simplifies to
\\
\begin{equation}
\int_{0.2}^{0.8} D_Q^K(z,{\rm Q}^2)dz  = 5\frac{dN^K(x,{\rm Q}^2)}{dN^{DIS}(x,{\rm Q}^2)}.
\label{eq4}
\end{equation}
\\
Eq.~(\ref{eq4}) has the property that 
the LHS is independent of $x$, while the RHS is potentially a
function of $x$. For the $x$ region in which this equation is valid, 
the RHS
must be independent of $x$. Interestingly, the HERMES data on
${dN^K}(x,{\rm Q}^2)/{dN^{DIS}}(x,{\rm Q}^2)$ are consistent 
with having a flat
$x$ dependence for $x > 0.1$, suggesting the validity of
Eq.~(\ref{eq4}) in the $x > 0.1$ region.
The HERMES collaboration obtained a linear fit to the
${dN^K}(x,{\rm Q}^2)/{dN^{DIS}}(x,{\rm Q}^2)$ data at $x>0.1$ 
as ${dN^K}(x,{\rm Q}^2)/{dN^{DIS}}(x,{\rm Q}^2)=(0.102\pm 0.002) 
+ (0.013 \pm 0.010)x$, 
shown as the curve
in Fig.~\ref{fig1}(a). This corresponds to a value of $
\int_{0.2}^{0.8} D_Q^K (z,{\rm Q}^2)dz 
= 5{dN^K}(x,{\rm Q}^2)/{dN^{DIS}}(x,{\rm Q}^2) = 0.514 \pm 0.010$ at 
${\rm Q}^2 = 2.5$ GeV$^2$ using Eq.~(\ref{eq4}) (note that at $x = 0.13$,
${\rm Q}^2 \sim 2.5$ GeV$^2$ for the HERMES data). From this determination of $ \int_{0.2}^{0.8} D_Q^K (z,{\rm Q}^2)dz$,
together with $Q(x,{\rm Q}^2)$ from the CTEQ6L PDF~\cite{CTEQ6L} and
$D_S^K(z,{\rm Q}^2)$ from the DSS FF~\cite{DSS}, $S(x,{\rm Q}^2)$ can
be readily obtained using Eq.~(\ref{eq2}). The slight ${\rm
  Q}^2$-dependence for $D_Q^K (z,{\rm Q}^2)$ was taken into account by
the same scale dependence as the DSS FF. The extracted $S(x)$ at ${\rm
  Q}^2 = 2.5$ GeV$^2$ with statistical errors only, shown as
``HERMES2014-set2'' in Fig.~\ref{fig1}(b), largely vanishes for
$x>0.1$, which reflects the assumption used in this approach. In
particular, Eqs.~(\ref{eq3}) and~(\ref{eq4}) ensure that $S(x) \to 0$
for $x>0.1$, where the validity of Eq.~(\ref{eq4}) is assumed by
HERMES~\cite{hermes14}. This striking result of vanishing $S(x)$ at
$x>0.1$ is at variance with all existing PDFs, including CTEQ6L and
CTEQ6.5S-0 shown in Fig.~\ref{fig1}(b).

The method of extracting $\int_{0.2}^{0.8} D_Q^K (z,{\rm Q}^2)dz$ from
the HERMES kaon multiplicity data is interesting, but it does not take
into account the constraints provided by the bulk of existing $e^+
e^-$ and SIDIS data. As discussed earlier, a straightforward approach
to extract $S(x)$ from the HERMES data is to use the DSS kaon
fragmentation functions of $D_Q^K (z,{\rm Q}^2)$ and $D_S^K (z,{\rm
  Q}^2)$ from the latest global fit to extensive data
sets~\cite{DSS}. Note that the DSS FF gives $ \int_{0.2}^{0.8} D_Q^K
(z,{\rm Q}^2)dz = 0.435$ at ${\rm Q}^2 = 2.5$ GeV$^2$, which is
$\sim$20\% smaller than the HERMES result. Adopting this value from
the DSS FF in Eq.~(\ref{eq2}), there is room for non-zero values of
$S(x)$ at $x>0.1$, since $5(dN^K(x,{\rm Q}^2)/dN^{DIS}(x,{\rm Q}^2))$
is now greater than $\int_{0.2}^{0.8} D_Q^K (z,{\rm Q}^2)dz$. This
would also lead to larger values of $xS(x)$ in the small-$x$ region,
as shown as ``HERMES2014-set2'' (statistic errors only) in
Fig.~\ref{fig1}(b) (square points).


It is important to note that a flat $x$ dependence for ${dN^K}(x,{\rm
  Q}^2)/{dN^{DIS}}(x,{\rm Q}^2)$ data at $x>0.1$ does not ensure the
validity of Eq.~(\ref{eq4}). Indeed, it is plausible that the second
term on the RHS of Eq.~(\ref{eq3}) can contribute on the order of
several percents relative to the first term. This would lower the
value of $\int_{0.2}^{0.8} D_Q^K (z,{\rm Q}^2)dz$ and the extracted
$S(x)$ would be larger than the ``HERMES2014-set2" result in Fig. 1(b). To assess
the relative importance of the second term on the RHS of
Eq.~(\ref{eq3}), we have investigated the contribution of this term to
Eq.~(\ref{eq3}) using a variety of PDFs. We found that the ratios of
the second term to the first term on the RHS of Eq.~(\ref{eq3}) at $x
= 0.25$, which is near the mean $x$ value of the HERMES data at the $x
> 0.1$ region, are 6.2\%, 3.5\% and 4.4\% for CTEQ6L~\cite{CTEQ6L},
NNPDF2.3L~\cite{NNPDF2.3} and MMHT2014L~\cite{MMTH2014}
respectively. We extracted the values of $xS(x)$ from the HERMES kaon
multiplicity data following the same procedure taking into account the
finiteness of the second term on the RHS of Eq.~(\ref{eq3}). The
results by setting the ratios of the second term to the first term on
the RHS of Eq.~(\ref{eq3}) to be the average value 4.8\% are shown as
``HERMES2014-set3'' (statistical errors only) in
Fig.~\ref{fig1}(b) (triangular points). We note that the problems
encountered by the other two approaches, namely the large $S(x)$
  content at $x> 0.2$ in HERMES2014-set1, or the vanishing strange-quark
  content at $x > 0.1$ in HERMES2014-set2, are largely mitigated.


In the rest of this paper, we present the results on the extraction of
the intrinsic strange and non-strange sea by using the two results of
$S(x)$ obtained by HERMES~\cite{hermes14} together with the
``HERMES2014-set3'' result. For comparison, we also include the result
using the values of $S(x)$ from the earlier HERMES
publication~\cite{hermes08}. The goal of this study is to assess the
range of uncertainty in the extraction of the light-quark intrinsic
sea, resulting from the uncertainty in $S(x)$ extracted from the
HERMES data. Table~\ref{tab0} summarizes the label, extraction
  method and reference information for each $S(x)$ data set to be
  studied.

\begin{table*}[htb]
\begin{center}
\begin{tabular}{l|l|l}\hline
\noalign{\smallskip} 
Label & Method & Reference \\
\noalign{\smallskip}
\hline
\hline
\noalign{\smallskip}
HERMES2008      & Eq.~\ref{eq2} and Eq.~\ref{eq4} & Ref.~\cite{hermes08} \\
HERMES2014-set1 & Eq.~\ref{eq2} and DSS FF & Ref.~\cite{hermes14}, Fig.~2 \\
HERMES2014-set2 & Eq.~\ref{eq2} and Eq.~\ref{eq4} & Ref.~\cite{hermes14}, Fig.~4 \\
HERMES2014-set3 & Eq.~\ref{eq2} and Eq.~\ref{eq3} &  \\
\hline
\hline
\end{tabular}
\end{center}
\caption{The label, extraction method and reference for each $S(x)$
  data set.}
\label{tab0}       
\end{table*}

\section{Extraction of intrinsic light-quark seas}

In the BHPS model~\cite{brodsky80}, the probability of the $|u u d
\mathcal{Q} \bar{\mathcal{Q}} \rangle$ proton five-quark Fock state,
where quark $i$ carries a momentum fraction $x_i$, is given as
\begin{equation}
P(x_1, ...,x_5)=N_5\delta(1-\sum_{i=1}^5x_i)[m_p^2-\sum_{i=1}^5\frac{m_i^2}{x_i}
]^{-2},
\label{eq:prob5q_a}
\end{equation}
\noindent where the delta function ensures momentum
conservation. $N_5$ is the normalization factor, and $m_i$ is the mass
of quark $i$. The last two quarks ($i=4,5$) refer to intrinsic
  sea quark pair $\mathcal{Q} \bar{\mathcal{Q}}$ in the five-quark
  Fock state. The momentum distribution, $P(x_i)$, for quark $i$ is
obtained by integrating Eq.~(\ref{eq:prob5q_a}) over the momentum
fractions of the remaining quarks. An analytical expression for the
probability distribution $P(x_5)$ for $\bar{\mathcal{Q}}$ is
available~\cite{brodsky80} in the limit of $m_{4,5} >> m_p,
m_{1,2,3}$.  When $\mathcal{Q}$ is the lighter $u$, $d$, or $s$ quark,
for which one could no longer assume a large mass, we developed the
algorithm to calculate $P(x_5)$ according to Eq.~(\ref{eq:prob5q_a})
with Monte-Carlo techniques~\cite{chang11}, and $m_u = m_d =
  0.3$ GeV/$c^2$, $m_s = 0.5$ GeV/$c^2$, and $m_p = 0.938$ GeV/$c^2$.

The challenge for identifying the intrinsic seas is to separate them
from the much more abundant extrinsic seas. Two approaches were
considered~\cite{chang11}. The first is to select experimental
observables which have little or no contributions from the extrinsic
seas. The $\bar d(x) - \bar u(x)$, which was measured in a
  Drell-Yan experiment~\cite{E866}, is an example of
  flavor-nonsinglet quantities which are largely free from the
contributions of the extrinsic sea quarks, since the perturbative $g
\to \mathcal{Q} \bar{\mathcal{Q}}$ processes will generate $u \bar u$
and $d \bar d$ pairs with very similar probabilities and have little
or no contribution to this quantity. Another example for the
quantities largely free from the extrinsic sea is the SU(3) flavor-nonsinglet $\bar u(x) + \bar d(x) - s(x) - \bar s(x)$
  distribution, which is obtained from the HERMES data on $s(x) + \bar
  s(x)$ together with $\bar u(x) + \bar d(x)$ from PDF global
  analysis.

The second approach is to rely on their different $x$ dependencies. As
mentioned earlier, the extrinsic sea is more abundant in the small-$x$
region while the intrinsic sea is valence-like and is dominant in the
large-$x$ region. The HERMES $S(x)$ data in Ref.~\cite{hermes08}
showed an intriguing feature of a sharp rise towards small $x$
($x<0.1$) and a broad structure in the larger $x$ region. This
suggests the presence of two distinct components of the strange sea,
an extrinsic part dominating at small $x$ and an intrinsic component
in the $x > 0.1$ region. A comparison between the HERMES data and the
calculations using the BHPS model showed good
agreement~\cite{chang11}, supporting the interpretation that the data
at $x>0.1$ have a significant contribution from the intrinsic sea.

\begin{figure}[htbp]
\includegraphics[width=0.9\textwidth]{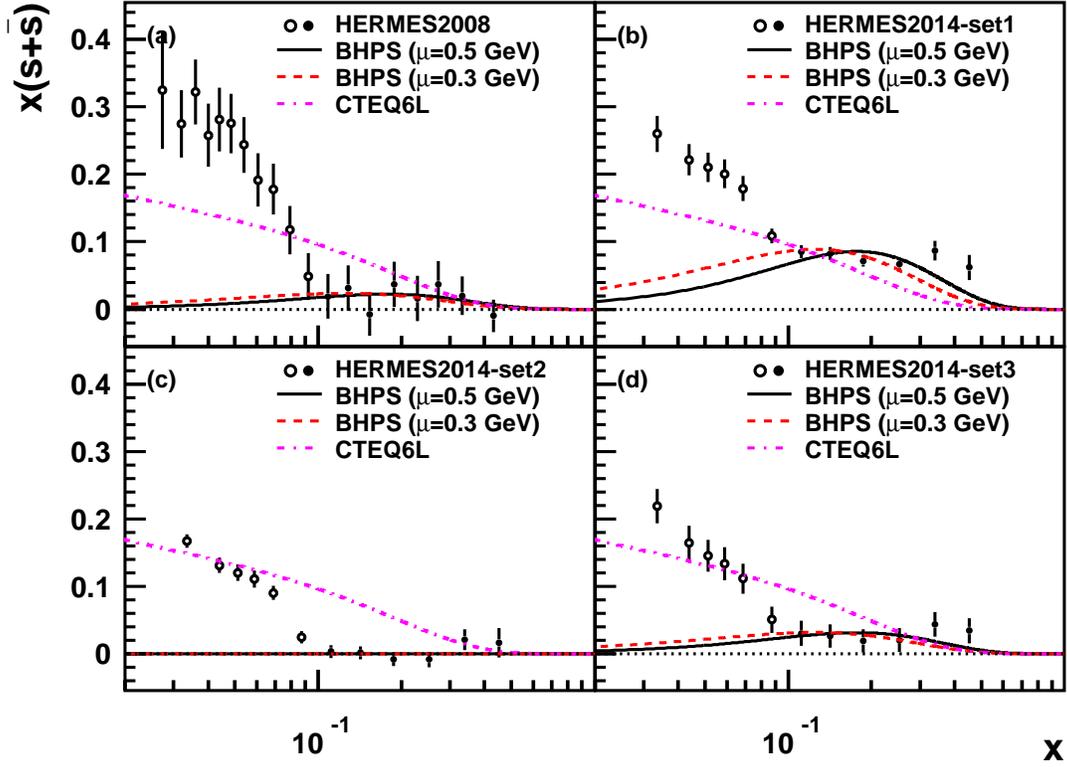}
\caption{Comparison of the HERMES $x(s(x) + \bar s(x))$ data with the
  calculations based on the BHPS model. The solid black and dashed red
  curves are obtained by evolving the BHPS result to ${\rm Q}^2 = 2.5$
  GeV$^2$ using the initial scale $\mu = 0.5$ GeV and $\mu = 0.3$ GeV,
  respectively. The normalizations of the calculations are adjusted to
  fit the data at $x > 0.1$, denoted by solid circles. The blue
  dash-dot and dotted lines denote the $x(s(x) + \bar s(x))$ from
  CTEQ6L~\cite{CTEQ6L} and CTEQ6.5S-0~\cite{CTEQ6.5s},
  respectively. The labels (a), (b), (c) and (d) denote the different
  inputs of $xS(x)$ from HERMES2008, HERMES2014-set1, HERMES2014-set2
  and HERMES2014-set3.}
\label{fig2}
\end{figure}

The moment of $P(x_5)$ is defined as ${\cal P}^{\mathcal{Q}
    \bar{\mathcal{Q}}}_5$ ($\equiv \int^{1}_{0} P(x_5) d x_5$) and
  represents the probability of the $|uud \mathcal{Q}
  \bar{\mathcal{Q}} \rangle$ five-quark Fock state in the proton. We
take the same approach as described in Ref.~\cite{chang11} to extract
the five-quark components of the proton, ${\cal P}_5^{u \bar u}$,
${\cal P}_5^{d \bar d}$ and ${\cal P}_5^{s \bar s}$. First, the
difference ${\cal P}_5^{d \bar d} - {\cal P}_5^{u \bar u}$ was
constrained to be $0.118 \pm 0.012$ by the normalization of $\bar d(x)
- \bar u(x)$ from the measurement of Fermilab E866 Drell-Yan
experiment~\cite{E866}. The ${\cal P}_5^{s \bar s}$ is obtained from
four different sets of data for $xS(x)$ at $x>0.1$ and ${\rm Q}^2 =
2.5$ GeV$^2$: HERMES2008, HERMES2014-set1, HERMES2014-set2, and
  HERMES2014-set3, respectively. The total errors of $xS(x)$
  obtained by the square-root sum of the statistical and systematic
  errors are used in the analysis. Figure~\ref{fig2} shows the fit to
$xS(x)$ at $x>0.1$ using the BHPS model to extract the intrinsic sea
for these four sets of data. The solid and dashed curves are obtained
by evolving the BHPS result to ${\rm Q}^2 = 2.5$ GeV$^2$ using the
initial scale value of $\mu = 0.5$ GeV and $\mu = 0.3$ GeV,
respectively. The normalization of the calculations are adjusted to
fit the data at $x > 0.1$. The $xS(x)$ from CTEQ6L~\cite{CTEQ6L} and
CTEQ6.5S-0~\cite{CTEQ6.5s} are also shown.

\begin{figure}[htb]
\includegraphics[width=0.9\textwidth]{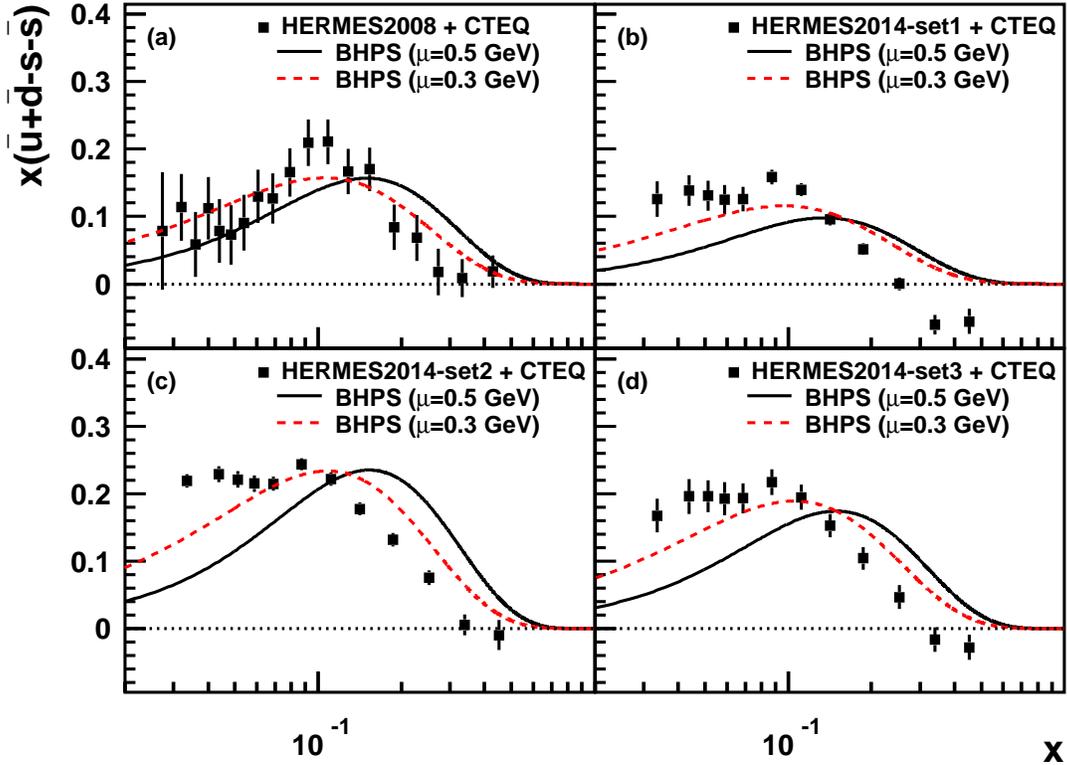}
\caption{Comparison of the $x(\bar u(x) + \bar d(x) - s(x) - \bar
  s(x))$ with the calculations based on the BHPS model. The solid
  black and dashed red curves are obtained by evolving the BHPS result
  to ${\rm Q}^2 = 2.5$ GeV$^2$ using $\mu = 0.5$ GeV and $\mu = 0.3$
  GeV, respectively. The normalizations of the calculations are
  adjusted to fit the data. The labels (a), (b), (c) and (d) denote
    the different inputs of $xS(x)$ from HERMES2008, HERMES2014-set1, HERMES-set2 and HERMES-set3.}
\label{fig3}
\end{figure}

Combining the HERMES data on $x(s(x)+ \bar s(x))$ with the $x(\bar
d(x) + \bar u(x))$ distributions determined from the global analysis
of CTEQ6.6~\cite{CTEQ6.6}, the quantity $x(\bar u(x) + \bar d(x) -
s(x) - \bar s(x))$ can be obtained and compared with the calculation
of the intrinsic sea in the BHPS model for the determination of ${\cal
  P}_5^{u \bar u} + {\cal P}_5^{d \bar d} - 2{\cal P}_5^{s\bar
  s}$. Figure~\ref{fig3} shows the comparison of $x(\bar u(x) + \bar
d(x) - s(x) - \bar s(x))$ with the calculations based on the BHPS
model. The solid black and dashed red curves are obtained by evolving
the BHPS result to ${\rm Q}^2 = 2.5$ GeV$^2$ using $\mu = 0.5$ GeV and
$\mu = 0.3$ GeV, respectively.

Putting these three quantities together we can determine the
probabilities ${\cal P}_5^{u \bar u}$, ${\cal P}_5^{d \bar d}$ and
${\cal P}_5^{s \bar s}$ for the $|u u d u \bar u\rangle$, $|u u d d
\bar d\rangle$, and $|u u d s \bar s\rangle$ configurations
individually. The extracted ${\cal P}_5^{u \bar u}$, ${\cal P}_5^{d
  \bar d}$ and ${\cal P}_5^{s \bar s}$, from four sets of data
(HERMES2008, HERMES2014-set1, HERMES2014-set2 and HERMES2014-set3) are
listed in Table~\ref{tab1}. Since the $s + \bar s$ were extracted
using a LO analysis of the HERMES data, we have also performed
analyses using the LO CTEQ6.5S-0~\cite{CTEQ6.5s} and
CTEQ6L~\cite{CTEQ6L} PDFs in addition to the NLO
CTEQ6.6~\cite{CTEQ6.6} PDF. The results of ${\cal P}_5^{u \bar u}$ and
${\cal P}_5^{d \bar d}$ are shown in parentheses.
Table~\ref{tab1} shows that the extracted values of ${\cal P}_5^{s \bar s}$,
varying from zero to $\sim$0.11, depends sensitively on the choice 
of $S(x)$. It is interesting to note that the values of 
${\cal P}_5^{s \bar s}$ extracted~\cite{chang_peng} using the HERMES2008
data are closest to those obtained with the HERMES2014-set3.
Table~\ref{tab1} also shows that the values of ${\cal P}_5^{u \bar u}$ and
${\cal P}_5^{d \bar d}$ are relatively insensitive to the choice of 
$S(x)$. Finally, the results have only minor dependence on the
choice of the PDF.
\begin{table*}[htb]
\begin{center}
\begin{tabular}{lcccc}\hline
\noalign{\smallskip} 
$xS(x)$ & $\mu$ (GeV) & ${\cal P}_5^{u \bar u}$ & ${\cal P}_5^{d \bar d}$ & ${\cal P}_5^{s \bar s}$ \\
\noalign{\smallskip}
\hline
\hline
\noalign{\smallskip}
HERMES2008      & 0.5 & 0.120 (0.128, 0.112) & 0.238 (0.246, 0.230) & 0.022 \\
HERMES2008      & 0.3 & 0.161 (0.174, 0.145) & 0.279 (0.292, 0.263) & 0.029 \\
HERMES2014-set1 & 0.5 & 0.125 (0.131, 0.124) & 0.243 (0.249, 0.242) & 0.086 \\
HERMES2014-set1 & 0.3 & 0.194 (0.202, 0.188) & 0.312 (0.320, 0.306) & 0.111 \\
HERMES2014-set2 & 0.5 & 0.178 (0.187, 0.167) & 0.296 (0.305, 0.285) & 0.000 \\
HERMES2014-set2 & 0.3 & 0.229 (0.242, 0.211) & 0.347 (0.360, 0.329) & 0.000 \\
HERMES2014-set3 & 0.5 & 0.148 (0.156, 0.142) & 0.266 (0.274, 0.260) & 0.031 \\
HERMES2014-set3 & 0.3 & 0.213 (0.225, 0.200) & 0.331 (0.343, 0.318) & 0.039 \\
\hline
\hline
\end{tabular}
\end{center}
\caption{The extracted values of ${\cal P}_5^{u \bar u}$, ${\cal
    P}_5^{d \bar d}$ and ${\cal P}_5^{s \bar s}$ from
  E866~\cite{E866}, CTEQ6.6~\cite{CTEQ6.6} and four sets of HERMES's
  data (HERMES2008, HERMES2014-set1, HERMES2014-set2 and HERMES2014-set3) assuming two
  initial scales ($\mu$) for the BHPS five-quark distributions. The
  results of ${\cal P}_5^{u \bar u}$ and ${\cal P}_5^{d \bar d}$ using
  CTEQ6.5S-0~\cite{CTEQ6.5s} and CTEQ6L~\cite{CTEQ6L} PDFs are also
  shown in parentheses.}
\label{tab1}       
\end{table*}

\section{Discussion}

\begin{figure}[htb]
\centering
\includegraphics[width=0.9\textwidth]{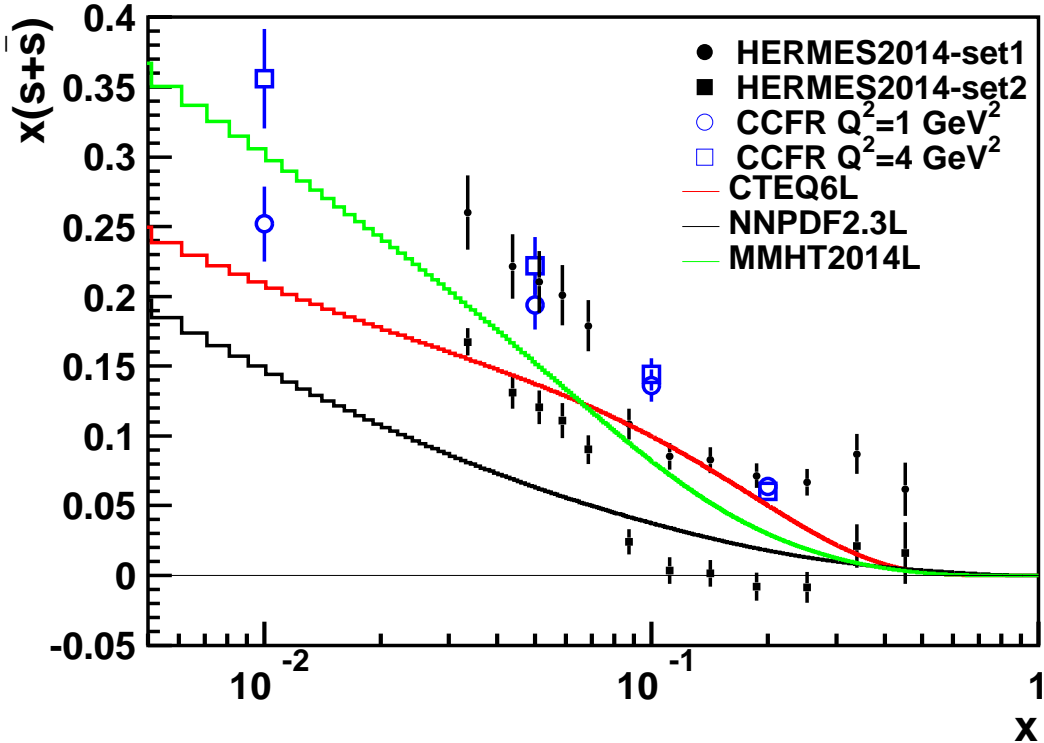}
\caption{The strange parton distribution $x(s + \bar s)$ from 
    HERMES2014-set1 and HERMES2014-set2 (statistical errors only) at ${\rm Q}^2
  = 2.5$ GeV$^2$, compared with those of CCFR (including statistical
  and systematic errors) at ${\rm Q}^2 = 1$ GeV$^2$ and $4$
  GeV$^2$~\cite{CCFR95}. The $x(s + \bar s)$ from three
    leading-order PDFs, CTEQ6L~\cite{CTEQ6L}, NNPDF2.3L~\cite{NNPDF2.3} 
and MMHT2014L~\cite{MMTH2014}, are overlaid.}
\label{fig4}
\end{figure}

The intrinsic five-quark strange component ${\cal P}_5^{s \bar s}$ in
the BHPS model is connected with the size of $xS(x)$ in valence-like,
i.e. large-$x$ region. The value of ${\cal P}_5^{s \bar s}$ is about
2-3\% for the HERMES2008 data~\cite{hermes08} and is either reduced to
less than 0.1\% or enhanced to 8-10\% depending on the choice of the
data sets. In Fig.~\ref{fig4} we compare the HERMES's SIDIS results of
$xS(x)$ distributions at ${\rm Q}^2 = 2.5$ GeV$^2$ with the CCFR's
results at ${\rm Q}^2 = 1$ GeV$^2$ and $4$ GeV$^2$~\cite{CCFR95}. The
distributions of $xS(x)$ at ${\rm Q}^2 = 2.5$ GeV$^2$ from
CTEQ6L~\cite{CTEQ6L}, NNPDF2.3L~\cite{NNPDF2.3} and
MMHT2014L~\cite{MMTH2014} leading-order PDFs are overlaid. The
assumption of vanishing strangeness for $x > 0.1$, adopted in the
recent HERMES's analysis, leads to results clearly at odds with the
data from the neutrino DIS experiment and the results of all
PDFs. Overall, the results using DSS FF (``HERMES2014-set1'') agree
best with the CCFR Data. Table~\ref{tab1} shows that the value of
$P^{s \bar s}_5$ is of the order of 0.03 to 0.11 from HERMES2014-set2
and HERMES2014-set3. A reliable extraction of $xS(x)$ and $P^{s \bar
  s}_5$ would require a more precise knowledge on the kaon
fragmentation functions~\cite{DSSerror}, and a new global fit taking
into account the recent HERMES~\cite{hermes13} and
COMPASS~\cite{COMPASS} kaon SIDIS data would be most valuable.

\begin{figure}[htb]
\centering
\includegraphics[width=0.9\textwidth]{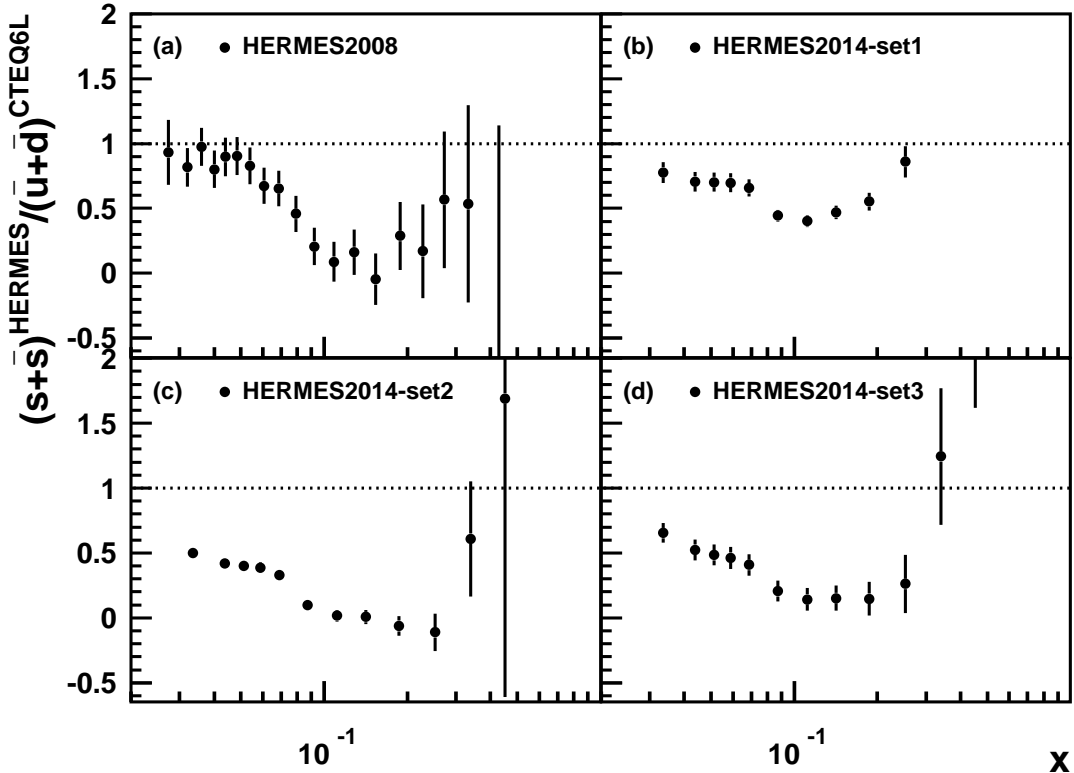}
\caption{The ratio of strange-to-nonstrange sea quarks $(s + \bar
  s)/(\bar u + \bar d)$ as a function of $x$. The $\bar u(x) + \bar
  d(x)$ is obtained from the CTEQ6L~\cite{CTEQ6L} PDF while the labels
  (a), (b), (c) and (d) denote the different input of $xS(x)$ from
  HERMES2008, HERMES2014-set1, HERMES2014-set2 and HERMES2014-set3,
  respectively.}
\label{fig5}
\end{figure}

Figure~\ref{fig5} shows the ratio of strange-to-nonstrange sea quarks
$(s + \bar s)/(\bar u + \bar d)$ as a function of $x$ using the HERMES data
of $(s + \bar s)$ and the $(\bar u + \bar d)$ from
CTEQ6L~\cite{CTEQ6L} at ${\rm Q}^2 = 2.5$ GeV$^2$. There are two
observations for the ratios: an enhancement at large $x$ and a rise
towards 1 at very small $x$. The first observation is consistent with
the existence of intrinsic strange sea which is distributed in larger
$x$ region relative to the intrinsic non-strange one because $m_s >
m_{u,d}$. The second observation suggests the presence of SU(3) flavor
symmetry in the small-$x$ region and is consistent with the
strange-to-down antiquark ratio $r_s = 1.00^{+0.25}_{-0.28}$ at
$x=0.023$ and ${\rm Q}^2 = 1.9$ GeV$^2$ from ATLAS~\cite{Atlas_WZ}.
This is also consistent with the expectation that the extrinsic sea,
which dominates at small $x$, is flavor independent.

\section{Summary}

In summary, we have studied the implications of the latest HERMES
  data on the strange-quark content and the intrinsic light-quark sea
  in the proton. We show that the striking result of a vanishing
  $xS(x)$ at $x>0.1$, reported as a favored solution by HERMES, is due
  to an assumption with a significant systematic uncertainty. We have
  calculated the five-quark components based on the BHPS model using
  the latest HERMES results on $xS(x)$. The new HERMES results affect
the strange content quantitatively but do not exclude 
the existence of intrinsic light-quark
component in nucleon sea. The $x$ dependence of the
strange-to-nonstrange sea quark ratio, $(s + \bar s)/(\bar u + \bar
d)$, is also in good qualitative agreement with the presence of both
the extrinsic and the intrinsic seas in the proton.  A reliable
extraction of $xS(x)$ and the intrinsic strange quark sea calls for a
more precise knowledge on the kaon fragmentation functions, and a new
global fit taking into account the recent kaon SIDIS data.

This work was supported in part by the National Science Council of the
Republic of China and the U.S. National Science Foundation.

\end{document}